\documentstyle[12pt]{article}

\def\llsymbol#1{\@llsymbol{\@nameuse{c@#1}}}
\def\@llsymbol#1{\ifcase#1\or {}\or {'}\or {''}\or {'''}\or
   {''''}\or {'''''}\or  \else\@ctrerr\fi\relax}
\newcounter{contador}
\newcommand{\letra}{
   \stepcounter{equation}
   \setcounter{contador}{\value{equation}}
   \setcounter{equation}{0}
   \renewcommand{\theequation}{\thecontador.\alph{equation}}}
\newcommand{\antiletra}{
   \renewcommand{\theequation}{\arabic{equation}}
   \setcounter{equation}{\value{contador}}}

\def\citet{\@ifnextchar [{\@tempswatrue\@citey}{\@tempswafalse\@citey[]}}

\def\@citey[#1]#2{\if@filesw\immediate\write\@auxout{\string\citation{#2}}\fi
  \def\@citec{}\@cite{\@for\@cited:=#2\do
    {\@citec\def\@citec{--}\@ifundefined
       {b@\@cited}{{\bf ?}\@warning
       {Citation `\@cited' on page \thepage \space undefined}}%
\hbox{\csname b@\@cited\endcsname}}}{#1}}
\newcommand{\natur}{{\kern+.19em\sf{I}\kern-.63em\sf{1}\kern+.83em\kern-.20em}}
\newcommand{\catur}{{\kern+.19em\sf{I}\kern-.43em\sf{C}\kern+.83em\kern-.20em}}

\begin{document}

\noindent
CBPF-NF-054/94\\[3mm]
DFTT-045/94

\begin{center}
{\Large\bf The Quantum Algebraic Structure of the Twisted XXZ Chain}

\ \\[1.5mm]

{\sl M.R-Monteiro$^{1,a}$,
I. Roditi$^{1,b}$, L.M.C.S. Rodrigues$^{1,c}$
and S. Sciuto$^{2,d}$\/} \\[0.2cm]

$^1$Centro Brasileiro de Pesquisas F\'{\i}sicas - CBPF\\[-1mm]
Rua Dr. Xavier Sigaud, 150\\[-1mm]
22290-180 \ Rio de Janeiro, RJ, Brasil\\[0.5cm]

$^2$Dipartimento di Fisica Teorica dell' Universit\'a di Torino
\\[-1mm] and
Sezione di Torino dell' INFN, Via P. Giuria 1, I-10125, Torino, Italy
\\[2cm]

{\sc Abstract}
\end{center}
We consider the Quantum Inverse Scattering Method with a new R-matrix
depending on two parameters $q$ and $t$. We find that the underlying
algebraic structure is the two-parameter deformed algebra
$SU_{q,t}(2)$ enlarged by introducing an element belonging to the centre. The
corresponding Hamiltonian describes the
spin-1/2 XXZ model with twisted periodic boundary conditions.


\noindent
\rule{7cm}{0.2mm}\\
{\footnotesize e-mail addresses: \\
(a) mmont@cbpfsu1.cat.cbpf.br \\[-1mm]
(b) roditi@cbpfsu1.cat.cbpf.br \\[-1mm]
(c) ligia@brlncc.bitnet \\[-1mm]
(d) sciuto@to.infn.it \\[-1mm]}


Many completely integrable one-dimensional quantum models have been
treated \cite{p} by the Quantum Inverse Scattering Method (QISM)\cite{e,ld}
which, among other achievements, led to the discovery of Quantum
Groups \cite{k} independently from Drinfeld and Jimbo\cite{v}. In this letter
we use this method to describe the relation between the spin-1/2 XXZ
chain with twisted periodic boundary conditions and the two-parameter
deformed algebra $SU_{q,t}(2)$ enlarged by introducing an element belonging to
the centre,
showing that the second parameter of
the deformation, $t$, is linked to the twist.

The QISM introduces an auxiliary  problem with the help of the
so-called Lax operator. In our model this operator is
\begin{equation}
L_n(\lambda ,t) = \left(
\begin{array}{cc}
t^{Z_n-S^3_n}\;sh[\gamma (\lambda +iS^3_n)] & iS^-_n\; sin\gamma \\
iS^+_nsin\gamma & t^{-Z_n-S_n^3}\; sh[\gamma (\lambda -iS_n^3)]
\end{array}
\right)\; ,
\end{equation}
where $\vec{S}_n$ and $Z_n$ are operators defined on the $n-th$ vectorial
space of the periodic $(\vec{S}_{N+1}\equiv \vec{S}_1, Z_{N+1}\equiv Z_1)$
chain, which in the fundamental
representation are given by
\begin{equation}
Z_n = \frac{1}{2}\;\natur_{\!\!\!\!n}\; , \vec{S}_n = \frac{1}{2}
\vec{\sigma}_n \; ,
\end{equation}
where $\vec{\sigma}$ are the Pauli matrices and $\natur$ is the
identity operator.

The $R$-matrix associated to the Lax operator (1) is
\begin{equation}
R(\lambda ,t) = \left(
\begin{array}{cccc}
a(\lambda ) & 0 & 0 & 0 \\
0 & c'(\lambda ) & b(\lambda ) & 0 \\
0 & b(\lambda ) & c''(\lambda ) & 0 \\
0 & 0 & 0 & a(\lambda )
\end{array}
\right)\; ,
\end{equation}
where
\begin{eqnarray}
&& a(\lambda ) = sh[\gamma (\lambda +i)] \nonumber \\
&& b(\lambda ) = isin\gamma \\
&& c'(\lambda ) = tc(\lambda )  \nonumber \\
&& c''(\lambda ) = t^{-1}c(\lambda ) \nonumber
\end{eqnarray}
and
\begin{equation}
c(\lambda ) = sh\; \gamma\lambda \; ;
\end{equation}
clearly, $R(\lambda ) = R(\lambda ,t=1)$ is the appropriate matrix for the
XXZ model \cite{li}. It is easy to check that the matrix $R(\lambda
,t)$ (eqs. (3-5)) satisfies the Yang-Baxter equation \cite{r,c}
\begin{equation}
R_{12}(\lambda_{12},t)R_{13}(\lambda_{13},t)R_{23}(\lambda_{23},t) =
R_{23}(\lambda_{23},t)R_{13}(\lambda_{13},t)R_{12}(\lambda_{12},t)
\end{equation}
and that $L_n(\lambda ,t)$ (eq. (1)) obeys the Fundamental Commutation
Relations $(FCR)$
\begin{equation}
R_{12}(\lambda_{12},t)L_n^1(\lambda_1,t)L_n^2(\lambda_2,t) =
L_n^2(\lambda_2,t)L_n^1(\lambda_1,t)R_{12}(\lambda_{12},t)\; .
\end{equation}
In (6) and (7), $\lambda_{ij}=\lambda_i-\lambda_j$ and
\begin{eqnarray}
&& R_{12} = \sum_i a_i\otimes b_i \otimes \natur \quad, \quad R_{13} = \sum_i
a_i\otimes
\natur \otimes b_i \; , \\
&&R_{23} = \sum_i \natur \otimes a_i \otimes b_i \; , \nonumber
\end{eqnarray}
with the $R(\lambda,t)$ matrix written as
\begin{equation}
R(\lambda ,t) = \sum_i a_i\otimes b_i
\end{equation}
and the upper indices in eq. (7) follow
\begin{equation}
L^1 = L\otimes \natur \; , \quad L^2 = \natur \otimes L.
\end{equation}
We notice that the $R$-matrix is defined on the tensor product of two
auxiliary spaces $\catur^{\!\!\!\!2}\otimes \catur^{\!\!\!\!2}$
and the $L$-matrix is defined on the tensor product of the auxiliary
space $\catur^{\!\!\!\!2}$ and the internal space
$\catur^{\!\!\!\!d}$, with $d$ the dimension of the representation of
the associated algebra satisfied by the operators in the elements of
the matrix $L$.

The reason for
$R(\lambda ,t)$ to satisfy eq. (6) is that it can be written
in terms of $R(\lambda )$ as\footnote{Equivalently one could
write
\[
g^1(g^2)^{-1} = t^{S^3\otimes Z-Z\otimes S^3}\; ,
\]
with $S^3$ and $Z$ given by eq. (2). This form is more appropriate if
one wishes to compare the algebraic structure  here presented with
ref. \cite{m}. In a forthcoming paper we shall discuss this
subject, as well as the relationship of our approach with the one in
ref. \cite{h}.}
\begin{equation}
R_{12}(\lambda ,t) = g^1(g^2)^{-1}R_{12}(\lambda )g^1(g^2)^{-1}\; ,
\end{equation}
where
\begin{equation}
g^1=g\otimes \natur \quad , \quad g^2=\natur \otimes g \quad ,
\quad g=t^{\frac{1}{2}S^3}
\end{equation}
and
\begin{equation}
[g^1g^2,R_{12}(\lambda )] = 0.
\end{equation}
Moreover, eq. (7) follows from eq. (6), because (for d=2)
\begin{equation}
L_n(\lambda ,t) = R_{o,n}(\lambda -\frac{i}{2},t),
\end{equation}
where ``o'' labels the auxiliary space. We also observe that
\begin{equation}
R(0,t) = P,
\end{equation}
where $P$ is the permutation operator on the tensor product of the
two spaces where the $R$-matrix is defined.

According to the standard procedure of the QISM, eq. (7) allows one to
build an infinite set of commuting operators
\begin{equation}
F(\lambda ,t) = Tr[L_N(\lambda ,t)\cdots L_2(\lambda ,t)L_1(\lambda
,t)]\; ,
\end{equation}
where both the matrix product and the trace are performed in the
auxiliary space.

In our case the Bethe Ansatz equations for the fundamental representation
are given by
\begin{equation}
\left(\frac{\alpha(\lambda_{\beta})}{\delta (\lambda_{\beta})}\right)^N =
t^{-N} \prod^M_{\atop{\alpha =1\atop \alpha \neq \beta}}
\left\{\frac{a(\lambda_{\beta}-\lambda_{\alpha})}
{a(\lambda_{\alpha}-\lambda_{\beta})}
\frac{c(\lambda_{\alpha}-\lambda_{\beta})}
{c(\lambda_{\beta}-\lambda_{\alpha})}\right\}\;
;\; \beta =1,\cdots ,M\leq N\; ,
\end{equation}
with
\begin{eqnarray}
&&\alpha (\lambda ) = sh\left[\gamma \left(\lambda +\frac{i}{2}\right)\right]
\\
&&\delta (\lambda ) = sh\left[\gamma \left(\lambda
-\frac{i}{2}\right)\right]\; ,
 \nonumber
\end{eqnarray}
explicitly showing the contribution due to the parameter $t$, as
$\alpha$, $\delta$, $a$ and $c$ are the same functions appearing in
the XXZ model.

In order to show the algebraic structure underlying the $R$ and $L$
matrices defined in eqs. (1) and (3-5) we perform a suitable similarity
transformation \cite{ld} on (6) and (7) which permits us to have the
following decomposition:
\begin{eqnarray}
&&\tilde{L}_n(\lambda ,t) = \frac{1}{2}
(e^{\lambda\gamma}L_+-e^{-\lambda\gamma}L_-) \\
&& \tilde{R}(\lambda_{ij},t) =
e^{\gamma\lambda_{ij}}R_{+}-e^{-\gamma\lambda_{ij}}R_-
\quad ; \nonumber
\end{eqnarray}
where
\begin{equation}
L_+ = \left(
\begin{array}{cc}
q^{S^3}t^{Z-S^3} & \Omega S^- \\
0 & q^{-S^3}t^{-Z-S^3}
\end{array}
\right)
\end{equation}

\[
L_- = \left(
\begin{array}{cc}
q^{-S^3}t^{Z-S^3} & 0 \\
-\Omega S^+ & q^{S^3}t^{-Z-S^3}
\end{array}
\right)
\]
and
\begin{equation}
R_+ =\left(
\begin{array}{cccc}
q & 0 & 0 & 0 \\
0 & t & \Omega & 0 \\
0 & 0 & t^{-1} & 0 \\
0 & 0 & 0 & q
\end{array}
\right)\quad ,
\end{equation}
with $\Omega = q-q^{-1}$ and $R_-=PR_+^{-1}P$.

Substituting eq. (19) in the Y-B equation (6) and in the FCR, eq.
(7), one gets the following independent equations:
\begin{eqnarray}
&&R_+L_{\varepsilon}^1L_{\varepsilon}^2 =
L_{\varepsilon}^2L_{\varepsilon}^1R_+ \qquad (\varepsilon = \pm 1) \\
&& R_+L_+^1L_-^2 = L_-^2L_+^1R_+ \; , \nonumber
\end{eqnarray}
which imply that the operators in the entries of $L$ must satisfy
\letra
\begin{eqnarray}
&&[S^3,Z] = [S^{\pm},Z] = 0 \\
&&[S^3,S^{\pm}] = \pm S_{\pm} \\
&&t^{-1}S^+S^--tS^-S^+=t^{-2S^3}[2S^3]_q\quad ,
\end{eqnarray}
\antiletra
where $[x]_q=(q^x-q^{-x})/(q-q^{-1})$ with $q=exp(i\gamma )$. Eqs.
(23) are the commutation relations of the two-parametric deformed SU(2)
\cite{m,s,b} with $Z$, an element of the center of the resulting algebra.
The coproduct
is obtained by considering the product of two $L_{\varepsilon}$ acting
on two internal spaces and we find:
\letra
\begin{eqnarray}
&&\Delta S^3 = S^3\otimes \natur + \natur \otimes S^3 \\
&&\Delta Z=Z\otimes \natur + \natur \otimes Z \\
&&\Delta S^{\pm} = q^{S^3}t^{\mp Z-S^3}\otimes S^{\pm}+ S^{\pm}\otimes q^{-S^3}
t^{\pm Z-S^3}.
\end{eqnarray}
\antiletra
The coproduct (24c) is related to the one in ref. \cite{b} by a
similarity transformation generated by the operator $t^{S^3\otimes
Z-Z\otimes S^3}$.

Following the QISM, a local Hamiltonian can be written as
\begin{equation}
H\ \propto \; \frac{\partial}{\partial \lambda} \ell g F(\lambda
,t)|_{\lambda = \frac{i}{2}}
\end{equation}
and thanks to eqs. (14-16), it becomes for the fundamental
representation of the algebra
\begin{eqnarray}
&&H = \sum^N_{i=1} \quad H_{i,i+1} \quad  (N+1\equiv 1) \\
&& H_{i,i+1} = \frac{J sin \gamma}{i\gamma}\; \frac{\partial}{\partial
\lambda} \hat{R}_{i,i+1}(\lambda ,t)|_{\lambda =0}\; , \nonumber
\end{eqnarray}
where $\hat{R}=PR$ and $R(\lambda ,t)$ is given by eqs. (3-5). In the
above equation $R_{i,i+1}(\lambda ,t)$ acts on the two internal
spaces $(i,i+1)$ instead of acting on two auxiliary spaces.

Substituting $R(\lambda ,t)$ given by eqs. (3-5) in eq. (26), apart from an
additive constant, we get
\begin{equation}
H = \frac{J}{2}
\sum^N_{i=1}\left[2t^{-1}\sigma_{i}^{+}\sigma_{i+1}^{-}+2t\sigma_i^-
\sigma_{i+1}^{+}+\frac{q+q^{-1}}{2} \sigma_i^z\sigma_{i+1}^z\right]\;
,
\end{equation}
where $\sigma^{\pm} = (\sigma^{x}\pm i\sigma^y)/2$.

Such a Hamiltonian is very similar to the XXZ model with
periodic boundary conditions but for each pair of sites $(i,i+1)$,
the site $(i+1)$ is rotated of an angle $\alpha\;\; (t=e^{i\alpha} )$ in
the $x-y$ plane with respect to the site $i$.

The similarity transformation generated by
$exp\{-i\frac{\alpha}{2}{\displaystyle{\sum^N_{\ell =1}}}(\ell
-1)\sigma_{\ell}^z\}$
takes the Hamiltonian (eq. (27)) to
\[
H = \frac{J}{2} \left[\sum^{N-1}_{n=1}\left(\sigma_n^x\sigma^x_{n+1}
+ \sigma_n^y\sigma_{n+1}^y + cos \gamma \;
\sigma_n^z\sigma_{n+1}^z\right) + \right.
\]
\begin{equation}
\left.  cos \gamma \; \sigma_N^z\sigma_1^z + 2t^{-N}\sigma_N^+
\sigma_1^- + 2t^N\sigma_N^- \sigma_1^+\right]\quad ,
\end{equation}
which is the well-known \cite{f} Hamiltonian for the XXZ chain with
twisted periodic boundary conditions.

It is amusing to observe that, thanks to eq. (13) and following the
procedure of ref. \cite{h}, the Hamiltonian (eq. (27)) could also be
obtained from the $R$-matrix of the XXZ model $R(\lambda )=R(\lambda
,t=1)$, using $L_n^{'}(\lambda ,t)=t^{S^3}L_n(\lambda ,t=1)\neq
L_n(\lambda ,t)$. Conversely, the untwisted XXZ model can be built from
$R(\lambda ,t)$ (eqs. (3-5)) and $L_n^{''}(\lambda
,t)=t^{-S^3}L_n(\lambda ,t)$. All these topics will be discussed in
detail in a forthcoming paper.

Finally, we would like to point out that by introducing a central element $Z$
which enlarges the $SU_{q,t}(2)$ algebra, we make appear the underlying
algebraic structure of the so-called twisted XXZ model.

\section*{Acknowledgements}

S. Sciuto thanks CBPF for its kind hospitality and CNPq for financial
support. I. Roditi thanks D. Altschuler for useful discussions.

\end{document}